\documentclass[%
 aip,
 amsmath,amssymb,
reprint, onecolumn 
]{revtex4-1}

\usepackage{graphicx}
\usepackage{dcolumn}
\usepackage{bm}
\usepackage{amsmath}

\usepackage[normalem]{ulem}

\usepackage[utf8]{inputenc}
\usepackage[T1]{fontenc}
\usepackage{mathptmx}
\usepackage{etoolbox}
\usepackage{color,colordvi}
\usepackage{subcaption}

\makeatletter
\def\@email#1#2{%
 \endgroup
 \patchcmd{\titleblock@produce}
  {\frontmatter@RRAPformat}
  {\frontmatter@RRAPformat{\produce@RRAP{*#1\href{mailto:#2}{#2}}}\frontmatter@RRAPformat}
  {}{}
}%
\makeatother
\begin{document}

\preprint{AIP/123-QED}


\title{Parametric processes in nonlinear structures with reflections: a transfer matrix method approach}

\author{Salvador Poveda-Hospital}
\email{salvador.poveda-hospital@polymtl.ca}
\affiliation{Departement de genie physique, \'Ecole polytechnique de Montr\'eal, Montr\'eal, QC, H3T 1J4, Canada}

\author{Tadeu Tassis}
\affiliation{Departement de genie physique, \'Ecole polytechnique de Montr\'eal, Montr\'eal, QC, H3T 1J4, Canada}

\author{Yves-Alain Peter}
\affiliation{Departement de genie physique, \'Ecole polytechnique de Montr\'eal, Montr\'eal, QC, H3T 1J4, Canada}

\author{Nicol\'as Quesada}
\affiliation{Departement de genie physique, \'Ecole polytechnique de Montr\'eal, Montr\'eal, QC, H3T 1J4, Canada}
\author{Martin Houde}
\affiliation{Departement de genie physique, \'Ecole polytechnique de Montr\'eal, Montr\'eal, QC, H3T 1J4, Canada}

\date{\today}

\begin{abstract}
The Transfer Matrix Method is a powerful numerical tool for simulating wave propagation in layered media. It has been widely applied in many fields, although its use is typically restricted to passive media. 
In this paper, we develop the transfer method to simulate optical nonlinear generation and give examples for processes such as difference frequency generation, spontaneous parametric down conversion, and four-wave mixing including generation of counter propagating waves. This framework enables accurate simulation of complex multilayer structures and resonant cavities, providing a versatile tool for designing nonlinear photonic devices.
\end{abstract}

\maketitle

\section{\label{sec:introduction}Introduction}
Nonlinear optics is a branch of optics that studies how light interacts with a medium \cite{armstrong1962interactions, franken1961generation}. Nonlinear effects occur when the polarization of the medium depends nonlinearly on the electric field, for example, 
a polarization proportional to the square of the field
leads to second-order nonlinear effects, characterized by the coefficient $\chi^{(2)}$, 
a polarization proportional to the cube leads to third-order nonlinear effects, characterized by $\chi^{(3)}$, and so on for higher orders. In this paper, we focus on nonlinear effects that involve energy transfer between light waves.
The second-order processes where new frequencies are generated are: second-harmonic generation (SHG), where two photons of the same frequency combine to form one at twice the frequency; sum-frequency generation (SFG), where two photons of different frequencies generate a photon at their sum frequency; difference-frequency generation (DFG), where the interaction produces a photon at the frequency difference; and spontaneous parametric down-conversion (SPDC), where a single high-energy photon splits into two lower-energy photons.
All these nonlinear processes are described by differential equations. Those differential equations can be solved analytically for simple systems, such as waveguides or Gaussian beams. However, for complex systems, such as systems with varying coefficients (refractive index, nonlinear susceptibility, light intensity), the analytical equations remain hard to solve or are unsolvable. Hence, we must rely on numerical methods.
Recently, there has been great interest in understanding the spectral response of nonlinear systems, particularly in quantum optics. Many quantum applications require carefully engineered single-photon sources, but their design is inherently complex. Developing tools that can accurately simulate such systems is essential, as they can significantly improve the quality and performance of photon sources \cite{rohde2005frequency}. 

The transfer matrix method (TMM) is widely used in various engineering applications
\cite{mackay2022transfer}, such as analyzing the transmission of energy in transmission lines, microwave filter designs, modeling multilayer optical coatings \cite{luce2022tmm}, wave propagation in photonic crystals \cite{lin2003lattice} and metamaterials, among others. 
The strength of TMM lies in its ability to represent each layer of a system with a matrix that relates incoming and outgoing fields. By sequentially multiplying these matrices, one can efficiently compute the overall system response without explicitly tracking infinite internal reflections. 
Note that the TMM is valid only in the continuous-wave regime, or when the temporal envelope of the pulse is much longer than the round-trip time of the resonant structures.
This makes TMM a powerful and elegant tool for analyzing complex linear systems,
meaning their response is directly proportional to the input. 
Hence, conventional TMM appears to be incompatible with the simulation of nonlinear optical processes. 
Previous attempts to use TMM for second-harmonic generation\cite{Albuquerque2011Transfer, li2015application, huang2018new} and parametric processes \cite{bethune1991optical}  fail to handle key challenges such as phase evolution, bidirectional propagation, and arbitrary material profiles. In contrast, the method proposed here introduces a nonlinear transfer matrix that accounts for amplitude, phase, bidirectional propagation, and arbitrary material profiles, enabling accurate and efficient simulation of advanced nonlinear photonic devices, and rigorously addresses processes such as DFG, SPDC, FWM, and SFWM while consistently treating both forward and backward propagating fields.

In this article, we show how to apply the TMM to simulate optical nonlinear generation processes. 
In Section~\ref{sec:Methodology}, we construct the nonlinear propagation transfer matrix step by step.
In Section~\ref{subsec:DFG}, we show how this method is used to simulate DFG.
In Section~\ref{subsec:SPDC}, we build a Fabry-Perot nonlinear transfer matrix to compare the simulation results with published experimental results for the SPDC process \cite{sorensen2025simple}.
In Section~\ref{subsec:FWM}, we derive the transfer matrix for FWM processes and simulate a Fabry-Pérot resonator with Bragg reflector mirrors and compare it with published experimental results \cite{xie2020chip}. 
Lastly, in Section~\ref{subsec:counterProp} we extend the transfer matrix to include the generation of photons in counter-propagating directions and compare it to published experimental results \cite{amores2025generation}.
All simulation codes used in this work are openly available at our GitHub repository \cite{poveda2025repository}.

\section{Methodology and Results} \label{sec:Methodology}

Difference-frequency generation (DFG) involves the interaction of three waves in a second-order nonlinear medium.
A strong pump wave interacts with a weaker signal wave inside a 
second-order nonlinear medium. This interaction produces a third wave, called the idler.
This motivates the notation $p$, $s$ and $i$ for pump, signal, and idler.
This process satisfies energy conservation $\omega_p = \omega_i + \omega_s$, 
while also requiring momentum conservation, expressed by the phase-matching condition $k_p = k_s + k_i$ . 
We will provide a walkthrough on how to obtain the transfer matrix of the DFG process. This matrix serves as the fundamental building block for simulating systems with varying coefficients, where matrices with different coefficients can be stacked to represent the overall system.
The coupled differential equations for DFG are \cite{boyd2008nonlinear}

\begin{subequations} \label{eq:coupled_eq_DFG}
\begin{align}
\frac{dA_p}{dz} &= \frac{2i \omega_p^2 d_{\text{eff}}}{k_p c^2} f_{psi} \, A_s A_i   e^{- i \Delta k z} \label{eq:coupled_eq_DFG_a}~, \\
\frac{dA_s}{dz} &= \frac{2i \omega_s^2 d_{\text{eff}}}{k_s c^2} f_{psi} \, A_p A_i^* e^{i \Delta k z} \label{eq:coupled_eq_DFG_b}~, \\
\frac{dA_i}{dz} &= \frac{2i \omega_i^2 d_{\text{eff}}}{k_i c^2} f_{psi} \, A_p  A_s^* e^{i \Delta k z} \label{eq:coupled_eq_DFG_c}~,
\end{align}
\end{subequations}
where $\omega$ is the angular frequency, $d_{\text{eff}}$ is the effective nonlinearity, $k$ is the propagation constant, $c$ the speed of light, $f_{psi}$ is the transverse modal field overlap, $A$ is the amplitude of the electric field, $z$ the spatial variable and \( \Delta k = k_p - k_s - k_i \). We use $p$, $s$, $i$ to denote pump, signal, and idler. 
The effective nonlinear coefficient $d_{\text{eff}}$
represents the projection of the second-order susceptibility tensor $\chi^{(2)}$ 
onto the polarization directions of the interacting waves, it depends on the crystal symmetry and on the polarizations of the pump, signal, and idler fields.
The modal overlap is defined as \cite{yang2008spontaneous}
\begin{align}
    f_{psi}=\frac{\displaystyle\iint F_p^*(x,y) F_s(x,y) F_i(x,y) \text{d}x\text{d}y}
{\displaystyle\sqrt{\Big(\iint|F_p(x,y)|^2\text{d}x\text{d}y\Big)\Big(\iint|F_s(x,y)|^2\text{d}x\text{d}y\Big)\Big(\iint|F_i(x,y)|^2\text{d}x\text{d}y\Big)}} ~,
\end{align}
where $F(x,y)$ is the normalized transverse modal field.

Firstly, we analyze the example in which the pump laser is undepleted, $\frac{dA_p}{dz} = 0 $. That is, the electric field amplitude is constant over $z$. The coupled differential equations become the following
\begin{subequations} 
\begin{align}
\frac{dA_s}{dz} &= \frac{2i \omega_s^2 d_{\text{eff}}}{k_s c^2} \bar{A}_p A_i^* e^{i \Delta k z}~, \\
\frac{dA_i}{dz} &= \frac{2i \omega_i^2 d_{\text{eff}}}{k_i c^2} \bar{A}_p  A_s^* e^{i \Delta k z}~,
\end{align}
\end{subequations} 
where now $\bar{A}_p$ is a constant. 
These differential equations can be solved analytically, allowing us to find the expression for the amplitudes at position $L+\Delta L$ in terms of those at position $L$
\begin{subequations} 
\begin{align}
    A_s({\scriptstyle L + \Delta L}) = 
    \left[ A_s({\scriptstyle L}) \left( \cosh{g\Delta L}-\frac{i\Delta k}{2g} \sinh{g\Delta L} \right) 
    + \frac{\kappa}{g} A_i^*({\scriptstyle L}) \sinh{g\Delta L} \right] e^{i\Delta k \Delta L/2} ~,\\
    A_i({\scriptstyle L + \Delta L}) = 
    \left[ A_i({\scriptstyle L}) \left( \cosh{g\Delta L}-\frac{i\Delta k}{2g} \sinh{g\Delta L} \right) 
    + \frac{\kappa}{g} A_s^*({\scriptstyle L}) \sinh{g\Delta L} \right] e^{i\Delta k \Delta L/2} ~,
\end{align}
\end{subequations} 
where 
\begin{align}
    \kappa = \frac{2i \omega_i \omega_s d_{\text{eff}}\bar{A}_p({\scriptstyle L})}{\sqrt{k_ik_s}c^2}
\end{align}
and 
\begin{align} \label{eq:g_simple}
    g = \left(\kappa\kappa^* -(\Delta k/2)^2 \right)^{1/2} ~.
\end{align}
In the TMM
it is necessary to account for both the amplitude and phase of the electric field. To do this, we include the fast propagation phase term.
In a linear system,
the electric field is written as, $E(z) = A(z) e^{ikz}$, and after propagating over a small distance the field becomes  $E(z+\Delta L) = E(z) e^{ik \Delta L}$.
The electric fields after propagating through the nonlinear region become
\begin{subequations} \label{eq:Es_Ei}
\begin{align}
    E_s({\scriptstyle L + \Delta L}) = 
    \left[ E_s({\scriptstyle L}) \left( \cosh{g\Delta L}-\frac{i\Delta k}{2g} \sinh{g\Delta L} \right) 
    + \frac{\kappa}{g} E_i^*({\scriptstyle L}) \sinh{g\Delta L} \right] e^{\Delta k \Delta L/2} e^{ik_s \Delta L} \\
    E_i({\scriptstyle L + \Delta L}) = 
    \left[ E_i({\scriptstyle L}) \left( \cosh{g\Delta L}-\frac{i\Delta k}{2g} \sinh{g\Delta L} \right) 
    + \frac{\kappa}{g} E_s^*({\scriptstyle L}) \sinh{g\Delta L} \right] e^{\Delta k \Delta L/2} e^{ik_i \Delta L} 
\end{align}
\end{subequations} 
where now $\kappa$ 
contains the pump electric field at position $L$ 
\begin{align}
    \kappa = \frac{2i \omega_i\omega_s d_{\text{eff}}E_p({\scriptstyle L})}{\sqrt{k_ik_s}c^2} ~.
\end{align}

\begin{figure}[h]
    \centering
    \includegraphics[width=0.4\linewidth]{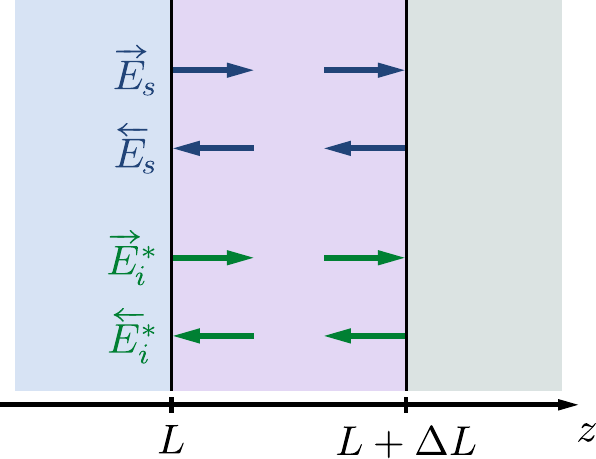}
    \caption{Schematic of nonlinear propagation single cell}
    \label{fig:FP_shcematic}
\end{figure}

To simulate any system, we need to include forward and backward propagating waves, annotated with a right upper arrow and a left upper arrow respectively.
To build the transfer matrix, we need to link the fields on the left $\overrightarrow{E}_{s}({\scriptstyle L})$, 
$ \overleftarrow{E}_{s}({\scriptstyle L}) $, 
$\overrightarrow{E}^\dagger_{i}({\scriptstyle L})$, 
$\overleftarrow{E}^\dagger_{i}({\scriptstyle L})$, 
in terms of the fields on the right 
$\overrightarrow{E}_{s}({\scriptstyle L + \Delta L})$, 
$ \overleftarrow{E}_{s}({\scriptstyle L + \Delta L})$, 
$\overrightarrow{E}^\dagger_{i}({\scriptstyle L + \Delta L})$, 
$\overleftarrow{E}^\dagger_{i}({\scriptstyle L + \Delta L})$, 
as shown in the schematic Fig.\ref{fig:FP_shcematic}. The steps to obtain the transfer matrix are as follows, first we introduce the constants $w_{11}^f$, $w_{22}^f$, $w_{21}^f$ and $w_{12}^f$ to simplify the notation.
\begin{subequations} 
\begin{align}
    \overrightarrow{E}_s({\scriptstyle L + \Delta L}) &=  \overrightarrow{E}_s({\scriptstyle L}) \, w_{11}^f +  \overrightarrow{E}_i^*({\scriptstyle L}) \, w_{12}^f \, ,  \\
    \overrightarrow{E}_i({\scriptstyle L + \Delta L}) &=  \overrightarrow{E}_i({\scriptstyle L}) \, w_{22}^f +  \overrightarrow{E}_s^*({\scriptstyle L}) \, w_{21}^f \, . 
\end{align}
\end{subequations} 
Secondly, we conjugate the idler equation and we find the left fields with respect to the right fields
\begin{subequations} 
\begin{align}
    \overrightarrow{E}_s(L)  &= \overrightarrow{E}_s(L+\Delta L) \frac{w_{22}^{f*}}{w_{22}^{f*}w_{11}^f-w_{21}^{f*}w_{12}^f} +  \overrightarrow{E}_i^*(L+\Delta L) \frac{-w_{12}^f}{w_{22}^{f*}w_{11}^f-w_{21}^{f*}w_{12}^f}, \\
    \overrightarrow{E}_i^*(L) &= \overrightarrow{E}_s(L+\Delta L) \frac{-w_{21}^{f*}}{w_{22}^{f*}w_{11}^f-w_{21}^{f*}w_{12}^f}  + \overrightarrow{E}_i^*(L+\Delta L) \frac{w_{11}^f}{w_{22}^{f*}w_{11}^f-w_{21}^{f*}w_{12}^f} ~.
\end{align}
\end{subequations} 
The final single cell nonlinear propagation transfer matrix is
\begin{align} \label{eq:TMM_fundamental_matrix}
    \mathbf{P}_{\text{NL}} =     \begin{pmatrix}
        \frac{w_{22}^{f*}}{w_{11}^fw_{22}^{f*}-w_{21}^{f*}w_{12}^f} & 0 & \frac{- w_{12}^f}{w_{11}^fw_{22}^{f*}-w_{21}^{f*}w_{12}^f} & 0\\
        0 & w_{11}^b  & 0 & w_{12}^b \\
        \frac{- w_{21}^{f*} }{w_{11}^fw_{22}^{f*}-w_{21}^{f*}w_{12}^f}   & 0 &\frac{w_{11}^f}{w_{11}^fw_{22}^{f*}-w_{21}^{f*}w_{12}^f} & 0\\
        0 & w_{21}^{b*} & 0 & w_{22}^{b*} 
    \end{pmatrix} ~,
\end{align}
where the upper index $f$ denotes the constants for the forward propagating wave and the upper index $b$ denotes the constants for the backward propagating wave. 
Since $\kappa$ now depends on the pump electric field propagating in both directions, we define a separate variable for each direction
\begin{subequations}
\begin{align}
    \overrightarrow{\kappa} &= \frac{2i \omega_i\omega_s d_{\text{eff}}\overrightarrow{E}_p\bigl(\scriptstyle L \bigr)}{\sqrt{k_ik_s}c^2} ~,\\
    \overleftarrow{\kappa} &= \frac{2i \omega_i\omega_s d_{\text{eff}}\overleftarrow{E}_p\bigl(\scriptstyle L + \Delta L \bigr)}{\sqrt{k_ik_s}c^2} ~.
\end{align}
\end{subequations}
Consequently, we define the variable $g$ for each propagation direction, as defined in Eq~\ref{eq:g_simple}.
The parameters in Eq.~\ref{eq:TMM_fundamental_matrix} are 
\begin{subequations} 
\begin{align}
    w_{11}^f &= \left( \cosh{\overrightarrow{g}\Delta L}-\frac{i\Delta k}{2\overrightarrow{g}} \sinh{\overrightarrow{g}\Delta L} \right) e^{i\Delta k \Delta L/2} e^{ik_s \Delta L} ~, \\
    w_{11}^b &= \left( \cosh{\overleftarrow{g}\Delta L}-\frac{i\Delta k}{2\overleftarrow{g}} \sinh{\overleftarrow{g}\Delta L} \right) e^{i\Delta k \Delta L/2} e^{ik_s \Delta L} ~, \\
    w_{12}^f &=  \frac{\overrightarrow{\kappa}}{\overrightarrow{g}} \sinh{(\overrightarrow{g}\Delta L)} \, e^{i\Delta k \Delta L/2} e^{ik_s \Delta L}~, \\
    w_{12}^b &=  \frac{\overleftarrow{\kappa}}{\overleftarrow{g}} \sinh{(\overleftarrow{g}\Delta L)} \, e^{i\Delta k \Delta L/2} e^{ik_s \Delta L}~, \\
    w_{22}^f &= \left( \cosh{\overrightarrow{g}\Delta L}-\frac{i\Delta k}{2\overrightarrow{g}} \sinh{\overrightarrow{g}\Delta L} \right) e^{i\Delta k \Delta L/2} e^{ik_i \Delta L} ~, \\
    w_{22}^b &= \left( \cosh{\overleftarrow{g}\Delta L}-\frac{i\Delta k}{2\overleftarrow{g}} \sinh{\overleftarrow{g}\Delta L} \right) e^{i\Delta k \Delta L/2} e^{ik_i \Delta L} ~, \\
    w_{21}^f &=  \frac{\overrightarrow{\kappa}}{\overrightarrow{g}} \sinh{(\overrightarrow{g}\Delta L)} \, e^{i\Delta k \Delta L/2} e^{ik_i \Delta L}~, \\
    w_{21}^b &=  \frac{\overleftarrow{\kappa}}{\overleftarrow{g}} \sinh{(\overleftarrow{g}\Delta L)} \, e^{i\Delta k \Delta L/2} e^{ik_i \Delta L}~. \\
\end{align}
\end{subequations} 
The fields on the left are now related by the fields on the right (of Fig.~\ref{fig:FP_shcematic}) by
\begin{align}
    \begin{pmatrix}
        \overrightarrow{E}_{s}({\scriptstyle L}) \\
        \overleftarrow{E}_{s}({\scriptstyle L}) \\
        \overrightarrow{E}^*_{i}({\scriptstyle L}) \\
        \overleftarrow{E}^*_{i}({\scriptstyle L})
    \end{pmatrix}  =
    \mathbf{P}_{\text{NL}}
    \begin{pmatrix}
        \overrightarrow{E}_{s}({\scriptstyle L + \Delta L}) \\
        \overleftarrow{E}_{s}({\scriptstyle L + \Delta L}) \\
        \overrightarrow{E}^*_{i}({\scriptstyle L + \Delta L}) \\
        \overleftarrow{E}^*_{i}({\scriptstyle L + \Delta L})
    \end{pmatrix}
    ~.
\end{align}

\subsection{Difference-frequency generation} \label{subsec:DFG}

\begin{figure}[h]
    \includegraphics[width=0.95\linewidth]{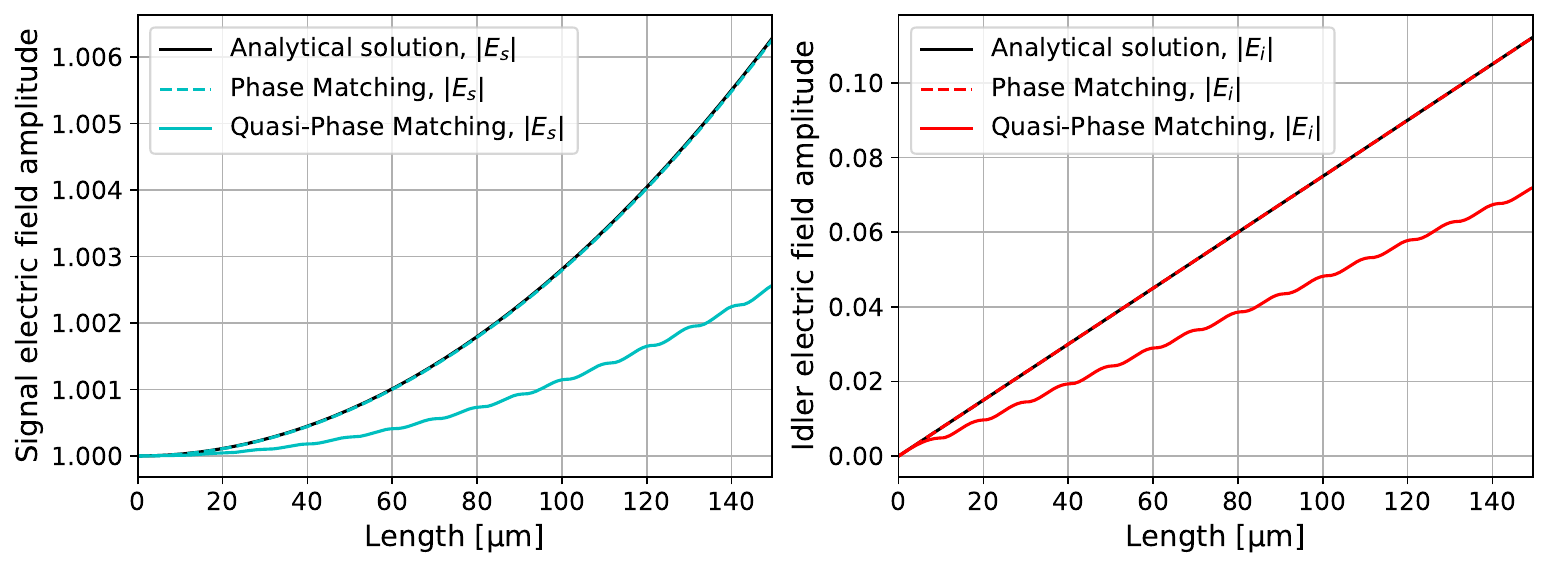} 
    \caption{TMM simulaton of DFG in straight waveguide with phase matching and quasi-phase matching, compared to the analytical solution. The signal input fields is 1~V/m and the idler input field is 0~V/m and the pump field is $10^7$~V/m.}
    \label{fig:wg_signal_idler}
\end{figure}

To validate the model, we first simulate propagation in a waveguide with and without phase-matching.
Phase matching describes the phase relation between the pump, signal, and idler, expressed as $\Delta k = k_p - k_i - k_s$.
If $\Delta k = 0$, we are phase matched and the nonlinear generation is optimized. However, in most cases the phase matching condition is not met and one must rely on Quasi-phase matching techniques.
Quasi-phase matching is obtained by the periodic variation of the nonlinear coefficient and modifies the phase mismatch such that 
\begin{align} \label{eq:quasiPM}
    \Delta k = k_p - k_i - k_s - 2\pi / \Lambda ~,
\end{align}
where $\Lambda$ is the coherence length. Depending on the material properties, this $ 2\pi / \Lambda$ adds an additional phase, which allows one to obtain $\Delta k = 0$.
To test the TMM method, we have built a stack of 150 matrices where each matrix represents the propagation through 1~\textmu m, for both cases (phase matched, quasi-phase matching). In the case of the quasi-phase matching the sign of $d_{\text{eff}}$ is flipped at every step when building the transfer matrix. 
Since this case has an analytical solution, we can confirm, as shown in Fig.~\ref{fig:wg_signal_idler}, that the results coincide perfectly with the analytical solution.
In Fig.~\ref{fig:wg_signal_idler}, the effect of phase matching is shown, with the periodic waves corresponding to the coherence length.

\subsection{Spontaneous parametric down-conversion} \label{subsec:SPDC}

In this subsection, we compare our simulation results with the experimental results reported by Sørensen et al. in Optics Express \cite{sorensen2025simple}. In Sørensen et al. paper, they propose a scattering matrix to simulate the amplitude of the electric fields inside a Fabry-Pérot cavity \cite{sorensen2025simple, kitaeva2004parametric}. Applying the TMM method leads to the same results, however, it has the advantage that propagation phases are taken into account and that the fields are organized in a way that it allows stacking of multiple matrices and hence being able to simulate more complex systems as nonlinear generation in a multi-layered media.

\begin{figure}[h]
    \centering
    \includegraphics[width=0.50\linewidth]{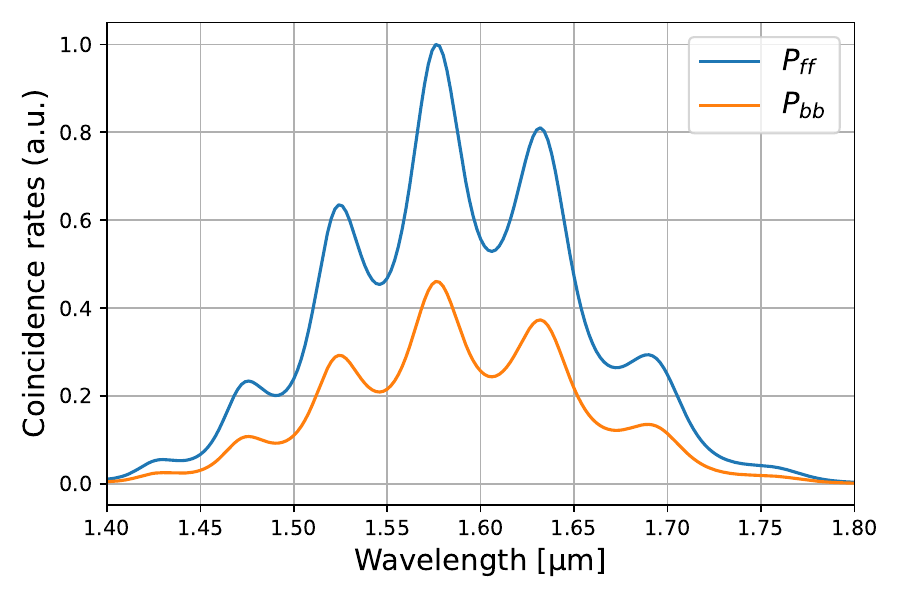}
    \caption{Fabry-Pérot TMM spectral results of the photon pairs in forward-forward emission (Eq.~\ref{eq:P_ff}) and backward-backward emission (Eq.~\ref{eq:P_bb}). }
    \label{fig:SPDC_TMM}
\end{figure}

In their experiments, the system is composed of 10.15~\textmu m of lithium niobate (LiNbO\textsubscript{3}) \cite{zelmon1997infrared} with air on one side and silicon (Si) on the other. This system is the simplest form of a Fabry-Pérot cavity, with a low quality factor since the reflection coefficients are low, $r_{air-LiNbO_3} = 0.36$ and $r_{LiNbO_3-Si} = 0.24$ . 
The system is pumped at 788~nm and single photons are detected on both sides. 
For this example, we define the transfer matrix of a partially reflective layer as
\begin{align}
    \mathbf{M} = 
    \begin{pmatrix}
             1/t_{s} & r_{s}/t_{s} & 0             & 0 \\
        r_{s}/t_{s} &      1/t_{s} & 0             & 0 \\
                    0 &             0 & 1/t_{i}      & r_{i}/t_{i} \\
                    0 &             0 & r_{i}/t_{i} &      1/t_{i} 
    \end{pmatrix} ~,    
\end{align}
where $r$ and $t$ are the Fresnel reflection and transmission coefficients. 
The transfer matrix describing this system is:
\begin{align} \label{eq:FP_SDPC_simple}
    &
    \begin{pmatrix}
        \overrightarrow{E}_{s}(0) \\
        \overleftarrow{E}_{s}(0) \\
        \overrightarrow{E}^\dagger_{i}(0) \\
        \overleftarrow{E}^\dagger_{i}(0)
    \end{pmatrix}  = 
    \mathbf{M}_{air-LiNbO_3} \,
    \mathbf{P}_{\text{NL}} \,
    \mathbf{M}_{LiNbO_3-Si}
    \begin{pmatrix}
        \overrightarrow{E}_{s}(L) \\
        \overleftarrow{E}_{s}(L) \\
        \overrightarrow{E}^\dagger_{i}(L) \\
        \overleftarrow{E}^\dagger_{i}(L)
    \end{pmatrix} ~.
\end{align}
Using the TMM formulation of Eq.~\ref{eq:FP_SDPC_simple} has the advantage of inherently accounting for all the infinite reflections within the resonant structure.
Because this approach is applied to SPDC, it is essential that the method preserve the bosonic commutation relations. 
To ensure this, we explicitly relate the classical fields in the TMM to the quantized field operators through the identifications ($\overrightarrow{E} \propto \overrightarrow{a} e^{ikz}$) and ($\overrightarrow{E}^{\dagger} \propto \overrightarrow{a}^\dagger e^{-ikz}$), where $\overrightarrow{a}$ is the annihilation operator for right-propagating modes and $\overrightarrow{a}^\dagger$ is the creation operator for left-propagating modes.
The analogous relations hold for the fields propagating to the left.
We verify this by re-expressing the transfer matrix in its scattering matrix form (from input to output).
The total transfer matrix is reordered so that the scattering matrix, $\bm{U}$, links the fields going inside the cavity and the field going out of the cavity
\begin{align}
    \begin{pmatrix}
        \overrightarrow{E}_{s}(L) \\
        \overleftarrow{E}_{s}(0) \\
        \overrightarrow{E}^\dagger_{i}(L) \\
        \overleftarrow{E}^\dagger_{i}(0)
    \end{pmatrix}  =
    \bm{U}
    \begin{pmatrix}
        \overrightarrow{E}_{s}(0) \\
        \overleftarrow{E}_{s}(L) \\
        \overrightarrow{E}^\dagger_{i}(0) \\
        \overleftarrow{E}^\dagger_{i}(L)
    \end{pmatrix}
    ~.
\end{align}
We then verify that matrix $\bm{U}$, obtained with the TMM simulation, preserves the Bogoliubov canonical commutation condition in the complex representation,
\begin{align} \label{eq:Bogoliubov_4x4}
\begin{aligned}
    U \, \Sigma \, U^\dagger = \Sigma ~, \\
    \text{with } ~ \Sigma = 
    \begin{pmatrix}
        I_2 & 0 \\
        0 & -I_2
    \end{pmatrix} ,
\end{aligned}
\end{align}
where $I_{2}$ is the $2\times 2$ identity matrix.
Because $U$ satisfies the Bogoliubov canonical commutation relation, it preserves the bosonic algebra of creation and annihilation operators. This guaranties that the output field operators can be expressed as linear combinations of the input operators via $U$, enabling the calculation of photon-emission probabilities directly from the elements of the scattering matrix.
The probability that two photons will be emitted forward is
\begin{align}
    \tilde{P}^{\mathrm{ff}} = 
\langle \text{vac} | \overrightarrow{a}_s^{\dagger}(L) \overrightarrow{a}_i^{\dagger}(L) \overrightarrow{a}_i(L) \overrightarrow{a}_s(L) | \text{vac} \rangle ~.
\end{align} 
where $|\text{vac}\rangle$ denotes the vacuum state and $\langle \text{vac}|$ its Hermitian conjugate.
Then $\tilde{P}^{\mathrm{ff}}$ can be calculated with the coefficients of the $\bm{U}$ matrix:
\begin{align} \label{eq:P_ff}
    \tilde{P}^{\mathrm{ff}} \propto |U_{20}|^2 \, (|U_{00}|^2 + |U_{02}|^2 + |U_{03}|^2) 
    + |U_{21}|^2  \, (|U_{01}|^2 + |U_{02}|^2 + |U_{03}|^2)  
    + 2 \mathrm{Re} \, (U_{00} \, U_{21} \, U_{20}^* \, U_{01}^*) ~.
\end{align}
The probability of two photons being emitted backward is
\begin{align} \label{eq:P_bb}
\begin{aligned}
    \tilde{P}^{\mathrm{bb}} &= 
    \langle \text{vac} | \overleftarrow{a}_s^{\dagger}(0) \overleftarrow{a}_i^{\dagger}(0) \overleftarrow{a}_i(0) \overleftarrow{a}_s(0) | \text{vac} \rangle \\
    &\propto 
      |U_{30}|^2 \, (|U_{10}|^2 + |U_{12}|^2 + |U_{13}|^2) 
    + |U_{31}|^2 \, (|U_{11}|^2 + |U_{12}|^2 + |U_{13}|^2)  
    + 2 \mathrm{Re} \, (U_{10} \, U_{31} \, U_{30}^* \, U_{11}^*) ~,
\end{aligned}
\end{align}
where the left-propagating operators are evaluated at $z=0$, corresponding to their final boundary condition.
The spectral response of the $\tilde{P}^{\mathrm{ff}}$ and $\tilde{P}^{\mathrm{bb}}$ corresponds to the experimental results, as shown in Fig.\ref{fig:SPDC_TMM}, where we can appreciate the influence of interference and confinement in the Fabry-Pérot cavity. 
In these simulations, we have included a Gaussian filter following the characteristics of the setup in Sørensen et al. \cite{sorensen2025simple}, and the results are in excellent agreement with their observations of the ratio between forward and backward-propagating coincidences.

\subsection{Nonlinear Optical Effects Beyond Phase Matching} \label{subsec:counterProp}

\begin{figure}[h]
    \centering
    \includegraphics[width=0.5\linewidth]{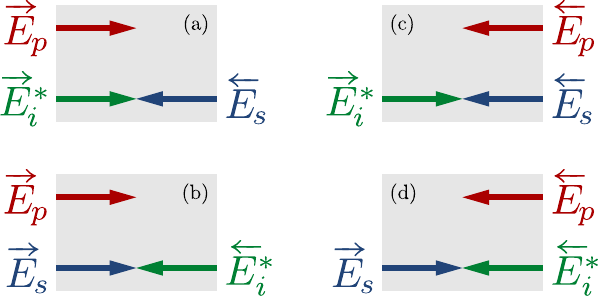}
    \caption{Schematic of the electric field nonlinear generation for counter-propagating idler and signal.}
    \label{fig:Beyond_Phase_Matching}
\end{figure}

Recent progress in nanostructured optical materials, plasmonics, and metasurfaces has made it possible to realize nonlinear optical processes without relying on phase matching \cite{kauranen2012nonlinear}. These strategies confine light to extremely small volumes and produce strong resonant field enhancements, leading to non-linear efficiencies that are far higher than those of the bulk materials themselves \cite{liu2016resonantly}. 
Fig.~\ref{fig:Beyond_Phase_Matching} shows a diagram of the additional electric fields that are taken into account when the phase matching is small compared to the size of the structure. It can be seen that this time there is nonlinear generation in one direction for the signal, while for the idler the direction is opposite. Similarly, the fields for the pump traveling to left can be plotted. This scheme still satisfies the condition of energy conservation.

The coupled differential equations for cases (a)--(d)
shown in Fig.~\ref{fig:Beyond_Phase_Matching}, are shown respectively in Appendix \ref{app:case_a}.
The single cell nonlinear propagation matrix is
\begin{align} \label{eq:TMM_contraProp}
    \begin{pmatrix}
        \overrightarrow{E}_{s}({\scriptstyle L}) \\
        \overleftarrow{E}_{s}({\scriptstyle L}) \\
        \overrightarrow{E}^\dagger_{i}({\scriptstyle L}) \\
        \overleftarrow{E}^\dagger_{i}({\scriptstyle L})
    \end{pmatrix}  =
    \begin{pmatrix}
    w_{11}^{f*}             &  0                       & - w_{12}^f        &-w_{14}^f-w_{14}^b \\
    0                       &  w_{11}^b                & w_{13}^b+w_{13}^f & w_{12}^b  \\
    - w_{21}^{f*}           & -w_{31}^{b*}-w_{31}^{f*} & w_{22}^{f}        & 0  \\
    w_{41}^{f*}+w_{41}^{b*} &  w_{21}^{b*}             & 0                 & w_{22}^{b*} 
    \end{pmatrix}
    \begin{pmatrix}
        \overrightarrow{E}_{s}({\scriptstyle L +\Delta L}) \\
        \overleftarrow{E}_{s}({\scriptstyle L +\Delta L}) \\
        \overrightarrow{E}^\dagger_{i}({\scriptstyle L +\Delta L}) \\
        \overleftarrow{E}^\dagger_{i}({\scriptstyle L +\Delta L})
    \end{pmatrix}
    ~,
\end{align}
where we have used an approximation that is valid when the simulation step is much smaller than the coherence length ($\Lambda$ in Eq.~\ref{eq:quasiPM}).
Eq.~\ref{eq:TMM_contraProp} is obtained by solving 6 ODEs
with our assumptions (undepleted pump, small length). The coefficients in Eq.~\ref{eq:TMM_contraProp} are

\begin{subequations} 
\begin{align}
    w_{11}^f &= e^{ik_s\Delta L}~, \\
    w_{11}^b &= e^{ik_s\Delta L}~, \\
    w_{22}^f &= e^{ik_i\Delta L}~, \\
    w_{22}^b &= e^{ik_i\Delta L}~, \\
    w_{13}^f &= -\frac{2i\omega_s^2 d_{\text{eff}}}{k_s c^2} \overrightarrow{E}_p({\scriptstyle L})  
    \Delta L \, \text{sinc}(\Delta k'' \, \Delta L /2) e^{-i\Delta k'' \, \Delta L/2}
    e^{i k_s \Delta L} ~, \\
    w_{31}^f &= \frac{2i\omega_i^2 d_{\text{eff}}}{k_i c^2} \overrightarrow{E}_p({\scriptstyle L}) 
    \Delta L \, \text{sinc}(\Delta k'' \, \Delta L /2) e^{i\Delta k'' \, \Delta L/2} 
    e^{i k_i \Delta L} ~, \\
    w_{14}^f &= \frac{2i\omega_s^2 d_{\text{eff}}}{k_s c^2} \overrightarrow{E}_p({\scriptstyle L}) 
    \Delta L \,  
    \text{sinc}(\Delta k' \, \Delta L /2) e^{i\Delta k' \, \Delta L/2}  
    e^{i k_s \Delta L} ~, \\
    w_{41}^f &= -\frac{2i\omega_i^2 d_{\text{eff}}}{k_i c^2} \overrightarrow{E}_p({\scriptstyle L}) 
    \Delta L \, 
    \text{sinc}(\Delta k' \, \Delta L /2) e^{-i\Delta k' \, \Delta L/2}   
    e^{ i k_i \Delta L} ~, \\
    w_{13}^b &= \frac{2i\omega_s^2 d_{\text{eff}}}{k_s c^2} \overleftarrow{E}_p ({\scriptstyle L+\Delta L}) 
    \Delta L \, \text{sinc}(\Delta k' \, \Delta L /2) e^{i\Delta k' \, \Delta L/2} 
    e^{i k_s \Delta L} ~, \\
    w_{31}^b &= -\frac{2i\omega_i^2 d_{\text{eff}}}{k_i c^2} \overleftarrow{E}_p({\scriptstyle L+\Delta L}) 
    \Delta L \, \text{sinc}(\Delta k' \, \Delta L /2) e^{-i\Delta k' \, \Delta L/2}  
    e^{i k_i \Delta L} ~, \\
    w_{14}^b &= -\frac{2i\omega_s^2 d_{\text{eff}}}{k_s c^2} \overleftarrow{E}_p({\scriptstyle L+\Delta L}) 
    \Delta L \, \text{sinc}(\Delta k'' \, \Delta L /2) e^{-i\Delta k'' \, \Delta L/2}   
    e^{i k_s \Delta L} ~, \\
    w_{41}^b &= \frac{2i\omega_i^2 d_{\text{eff}}}{k_i c^2} \overleftarrow{E}_p({\scriptstyle L+\Delta L}) 
    \Delta L \, \text{sinc}(\Delta k'' \, \Delta L /2) e^{i\Delta k'' \, \Delta L/2}  
    e^{i k_i \Delta L} ~.
\end{align}
\end{subequations} 
where $\Delta k' = k_p + k_i - k_s$ and $\Delta k'' = k_p - k_i + k_s$ are the phase mismatches for the new processes. The derivations on how to obtain the coefficients are  shown in Appendix \ref{app:case_a}. These phase-matching conditions are more stringent because the  coherence length is shorter.
Nevertheless, in nanostructured resonant devices with dimensions below the coherence length, phase-matching effects become negligible.

Now that we have derived the transfer matrix, we replicate the experimental results reported by A. P. Amores et al. in Physical Review A \cite{amores2025generation}. 
The system is composed of a periodically poled Rb-KTiOPO\textsubscript{4} crystal, where SPDC generates signal and idler photons in opposite directions at 1.596~\textmu m, pumped at 798~nm. 
The probability of a signal photon emitted forward and the idler backward, can be written in terms of the elements of the scattering matrix as \cite{sorensen2025simple}
\begin{align} \label{eq:P_fb}
    \tilde{P}^{\mathrm{fb}}  &= 
    \langle \text{vac} | \overrightarrow{a}_s^{\dagger}(L) \overleftarrow{a}_i^{\dagger}(0) \overleftarrow{a}_i(0) \overrightarrow{a}_s(L) | \text{vac} \rangle  \nonumber \\
    &\propto 
      |U_{30}|^2 \, (|U_{00}|^2 + |U_{02}|^2 + |U_{13}|^2) 
    + |U_{31}|^2 \, (|U_{02}|^2 + |U_{01}|^2 + |U_{03}|^2)  
    + 2 \mathrm{Re} \, (U_{30} \, U_{01} \, U_{00}^* \, U_{31}^*) ~. 
\end{align}
In this case the signal and idler are at the same frequency, so $\tilde{P}^{\mathrm{fb}}  = \tilde{P}^{\mathrm{bf}}$. 

\begin{figure}[h]
    \centering
    \includegraphics[width=0.49\linewidth]{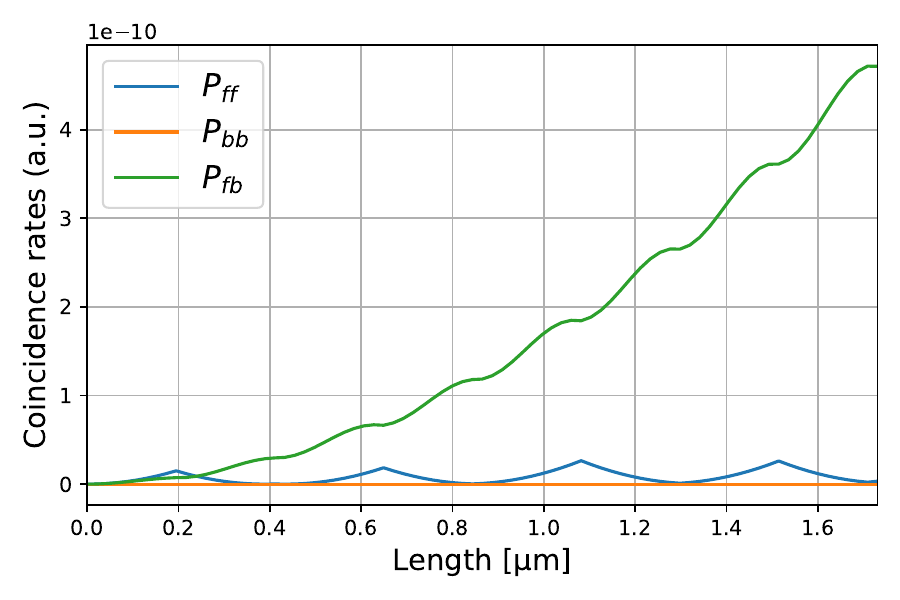}
    \includegraphics[width=0.49\linewidth]{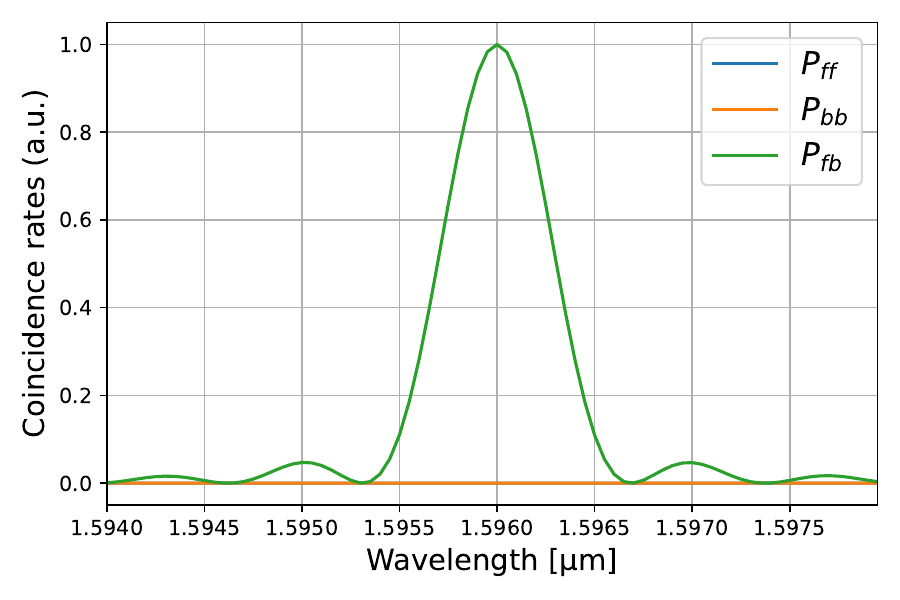}
    \caption{(a) Coincidence rate at different lengths of the crystal. (b) Spectral response for the forward-forward emitted probability, backward-backward emitted probability and forward-backward emitted probability, for a crystal length of 1~mm.}
    \label{fig:counter}
\end{figure}

As shown in Fig.~\ref{fig:counter} (a), for a periodic poling made for the counter propagating waves, the forward-backward emission increases, contrary to the forward-forward which oscillates around zero. 
These results are obtained with a crystal whose poling period is specifically engineered to achieve
quasi-phase matching for the counter-propagating waves, and this behavior is also confirmed experimentally \cite{amores2025generation}.
The backward-backward emission is zero since the pump is moving in the forward direction. Fig.~\ref{fig:counter} (b) shows the spectral response, which, as expected, follows a sinc-like function. 

\subsection{Four-Wave Mixing} \label{subsec:FWM}

In this section we study the Four-Wave Mixing (FWM) process where two identical photons generate signal and idler photons, satisfying energy conservation: $2  \omega_p = \omega_i + \omega_s$, 
while also requiring momentum conservation, expressed by the phase-matching condition $2 k_p = k_s + k_i$. 
This process is behind many modern photonics applications, such as frequency comb generation \cite{sefler1998frequency}, supercontinuum generation \cite{coen2002supercontinuum}, quantum photon-pair generation \cite{zhang2021squeezed}, among others.
The FWM equation of motion \cite{nonlinear-fiber-optics} are
\begin{subequations} 
\begin{align} \label{eq:coupled_eq_FWM}
\frac{dA_p}{dz} &= \frac{2 i n_2 \omega_p}{c} f_{psi} \, A_p^* A_s A_i e^{- i \Delta k z} ~, \\
\frac{dA_s}{dz} &= \frac{2 i n_2 \omega_s}{c} f_{psi} \, A_p^2 A_i^* e^{i \Delta k z}~, \\
\frac{dA_i}{dz} &= \frac{2 i n_2 \omega_i}{c} f_{psi} \, A_p^2 A_s^* e^{i \Delta k z}~,
\end{align}
\end{subequations} 
where this time $\Delta k = 2\,k_p - k_s - k_i$, $n_2$ is  the nonlinear-index coefficient  and $f_{psi}$ is the overlap integral \cite{Stolen1982Parametric} defined as 
\begin{align}
    f_{psi}
= \frac{\displaystyle\iint \left[ F_p^*(x,y) \right]^2 \, F_s(x,y) \, F_i(x,y) \, \text{d}x\text{d}y}
{\displaystyle\sqrt{\Big(\iint|F_p(x,y)|^2\text{d}x\text{d}y\Big)^2\Big(\iint|F_s(x,y)|^2\text{d}x\text{d}y\Big)\Big(\iint|F_i(x,y)|^2\text{d}x\text{d}y\Big)}},
\end{align}
where $F(x,y)$ are the transverse modal fields.
The equations of motion are very similar to those of DFG, the only differences are the constant coefficients, the phase mismatch, and the fact that the pump amplitude is squared. 
Assuming the undepleted pump approximation,
the nonlinear propagation transfer matrix is of the same form as Eq.~\ref{eq:TMM_fundamental_matrix} , where now the matrix coefficients are
\begin{subequations} 
\begin{align}
    w^{f}_{11} &=  \left( \cosh{\overrightarrow{g_s}\Delta L}-\frac{i\Delta k}{2\overrightarrow{g_s}} \sinh{\overrightarrow{g_s}\Delta L} \right)  e^{i\Delta k \Delta L/2} e^{ik_s \Delta L} ~, \\  
    w^{f}_{12} &= \frac{\overrightarrow{\kappa_s}}{\overrightarrow{g_s}} \sinh{(\overrightarrow{g_s}\Delta L)} \, e^{i\Delta k \Delta L/2} e^{ik_s \Delta L} ~,  \\
    w^{f}_{22} &= \left( \cosh{\overrightarrow{g_i}\Delta L}-\frac{i\Delta k}{2\overrightarrow{g_i}} \sinh{\overrightarrow{g_i}\Delta L} \right)  e^{i\Delta k \Delta L/2} e^{ik_i \Delta L} ~, \\
    w^{f}_{21} &= \frac{\overrightarrow{\kappa_i}}{\overrightarrow{g_i}} \sinh{(\overrightarrow{g_i}\Delta L)} \, e^{i\Delta k \Delta L/2} e^{ik_i \Delta L} ~, \\
    w^{b}_{11} &=  \left( \cosh{\overleftarrow{g_s}\Delta L}-\frac{i\Delta k}{2\overleftarrow{g_s}} \sinh{\overleftarrow{g_s}\Delta L} \right)  e^{i\Delta k \Delta L/2} e^{ik_s \Delta L} ~, \\  
    w^{b}_{12} &= \frac{\overleftarrow{\kappa_s}}{\overleftarrow{g_s}} \sinh{(\overleftarrow{g_s}\Delta L)} \, e^{i\Delta k \Delta L/2} e^{ik_s \Delta L} ~,  \\
    w^{b}_{22} &= \left( \cosh{\overleftarrow{g_i}\Delta L}-\frac{i\Delta k}{2\overleftarrow{g_i}} \sinh{\overleftarrow{g_i}\Delta L} \right)  e^{i\Delta k \Delta L/2} e^{ik_i \Delta L} ~, \\
    w^{b}_{21} &= \frac{\overleftarrow{\kappa_i}}{\overleftarrow{g_i}} \sinh{(\overleftarrow{g_i}\Delta L)} \, e^{i\Delta k \Delta L/2} e^{ik_i \Delta L} ~.    
\end{align}
\end{subequations} 
with 
\begin{subequations}
\begin{align}
    \overrightarrow{\kappa}_j &= \frac{2 i n_2 \omega_j f_{psi}}{k_j c^2} \overrightarrow{E}_p^2({\scriptstyle L}) \, ~, \\
    \overleftarrow{\kappa}_j &= \frac{2 i n_2 \omega_j f_{psi}}{k_j c^2} \overleftarrow{E}_p^2({\scriptstyle L+\Delta L}) \, ~,
\end{align}
\end{subequations}
where $j \in \{p,s,i \}$ labels the pump, signal and idler,
and 
\begin{subequations}
\begin{align}
    \overrightarrow{g_j} &= \sqrt{\overrightarrow{\kappa_j} \overrightarrow{\kappa_j}^* - (\Delta k/2)^2} ~, \\
    \overleftarrow{g_j} &= \sqrt{\overleftarrow{\kappa_j} \overleftarrow{\kappa_j}^* - (\Delta k/2)^2} ~.
\end{align}
\end{subequations}

\begin{figure}[h]
    \centering
    \includegraphics[width=0.49\linewidth]{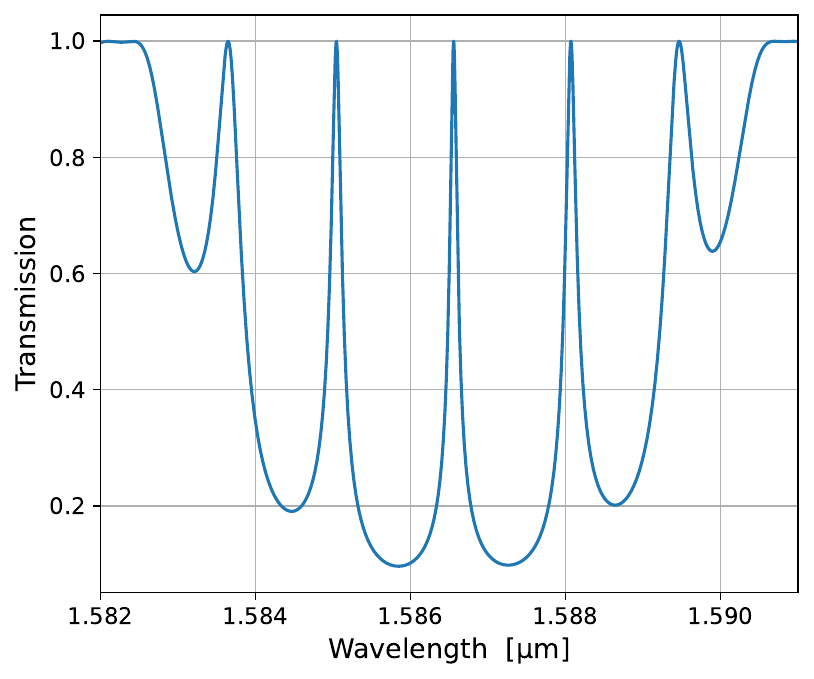}
    \includegraphics[width=0.49\linewidth]{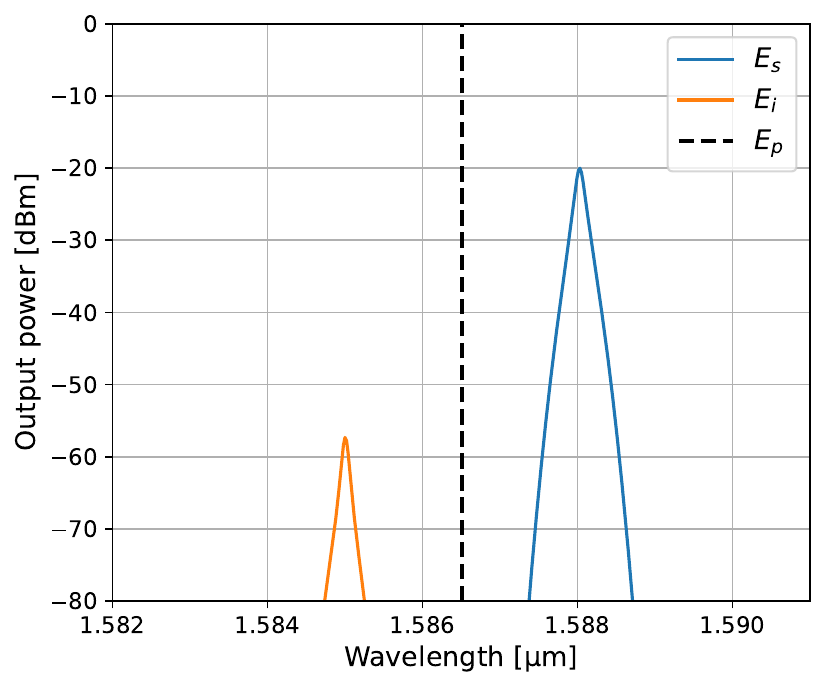}
    \caption{(a) Spectral thansmission of the Fabry-Pérot cavity. (b) Idler generation from stimulated FMW. Pump central frequency is at 1.5865~\textmu m}
    \label{fig:Transmission_Bragg}
\end{figure}

To validate the model, we compare our simulation results with the experimental data reported by Xie \textit{et al.}  \cite{xie2020chip}. The system consists of a Fabry–Pérot cavity formed by Bragg reflectors in a silicon nitride thin film. Figure~\ref{fig:Transmission_Bragg} shows the spectral transmission of the Farby-Pérot cavity.
Moreover, using Eq.~\ref{eq:Ep_at_z} depicted in Appendix~\ref{app:E_z}, we calculate the pump field intensity at every point of the Bragg reflectors.
To build the transfer matrix for this system, we concatenate a nonlinear propagation matrix with a mirror transfer matrix, so that the Bragg reflectors are also considered for the signal and idler. The transfer matrix describing this system is:
\begin{align} \label{eq:FP_SDPC_simple}
    &
    \begin{pmatrix}
        \overrightarrow{E}_{s}(0) \\
        \overleftarrow{E}_{s}(0) \\
        \overrightarrow{E}^\dagger_{i}(0) \\
        \overleftarrow{E}^\dagger_{i}(0)
    \end{pmatrix}  = 
    \left( \prod_{m=1}^{N}  \mathbf{P}_{\text{NL}}({\scriptstyle \Lambda_\text{BG}}) \, \mathbf{M}_{12} \, \mathbf{P}_{\text{NL}}({\scriptstyle \Lambda_\text{BG}}) \, \mathbf{M}_{21} \right) 
    \mathbf{P}_{\text{NL}}({\scriptstyle L_{\text{cavity}}})
    \left( \prod_{m=1}^{N}  \mathbf{P}_{\text{NL}}({\scriptstyle \Lambda_\text{BG}}) \, \mathbf{M}_{12} \, \mathbf{P}_{\text{NL}}({\scriptstyle \Lambda_\text{BG}}) \, \mathbf{M}_{21} \right) 
    \begin{pmatrix}
        \overrightarrow{E}_{s}({\scriptstyle 4N\Lambda_\text{BG}+ L_{\text{cavity}}}) \\
        \overleftarrow{E}_{s}({\scriptstyle 4N\Lambda_\text{BG}+ L_{\text{cavity}}})  \\
        \overrightarrow{E}^\dagger_{i}({\scriptstyle 4N\Lambda_\text{BG}+ L_{\text{cavity}}}) \\
        \overleftarrow{E}^\dagger_{i}({\scriptstyle 4N\Lambda_\text{BG}+ L_{\text{cavity}}})
    \end{pmatrix} ~.
\end{align}
where $L_{\text{cavity}}$ is the cavity length, $\Lambda_\text{BG}$ is half of the grating period, $N$ is the total number of gratings for each Bragg reflector, and $M_{12}$ and $M_{21}$ are the transfer matrices describing propagation from medium 1 to medium 2 and from medium 2 to medium 1, respectively, for each grating period.
In total a stack of 3200 matrices is used.
The cavity is pumped at the central peak (1.5865~\textmu m), a second laser, at 1.5885~\textmu m, provides the signal. At 1.5855~\textmu m, FWM nonlinear generation is observed. 
The simulations in Fig.~\ref{fig:Transmission_Bragg} closely reproduce the experimental data.

In addition, the transfer matrix of the entire system can be applied to calculate Spontaneous-FWM (SFWM). 
SFWM is widely used in integrated quantum photonics for heralded single-photon sources for applications in optical quantum computers \cite{PsiQuantum, helt2010spontaneous}.
The probability of two photons being emitted forward or backward, or one forward and the other one backward, can be calculated from the coefficients of the transfer matrix. Those equations are the same as for the SPDC process: Eq.\ref{eq:P_ff}, Eq.\ref{eq:P_bb} and  Eq.\ref{eq:P_fb}.

\section{Conclusion}

In conclusion, we have shown that the TMM method
yields the same results as the analytical solution of the differential coupled equations. We demonstrated that the TMM method can be applied to the SPDC process and that the transfer matrix is directly related to the photon-pair generation rate. The TMM method also corroborates the experimental results of SPDC obtained in Fabry–Pérot narrow cavities. 
We show that the TMM method is also applicable in FWM, replicating results in a Fabry-Pérot cavity made with Bragg Reflectors and its potential for SFWM.
Additionally, we have extended the transfer matrix to include 
counter propagating fields.
This method enables the simulation of new single-photon sources, the modeling of complex cavities, and, more generally, the simulation of structures where any parameter in the differential equations may vary (e.g., electric field amplitude, transverse modal field distribution, linear and nonlinear susceptibility). This constitutes a powerful tool that could be integrated into existing numerical frameworks. 
The TMM applied to nonlinear optics will support the progress in quantum photonic device design and classical nonlinear optics.

\begin{acknowledgments}
The authors would like to thank Régis Guertin and Marc-Antoine Bianki for helpful suggestions and discussions.
N. Q. and Y.-A. P. acknowledge support from the Ministère de l’Économie et de l’Innovation du Québec and the Natural Sciences and Engineering Research Council of Canada  Quantum Consortium Program (Quantamole).
\end{acknowledgments}

\section*{Data Availability Statement}

\appendix


\section{Field Distribution via the TMM Method} \label{app:E_z}

\begin{figure}[h]
    \centering
    \includegraphics[width=0.45\linewidth]{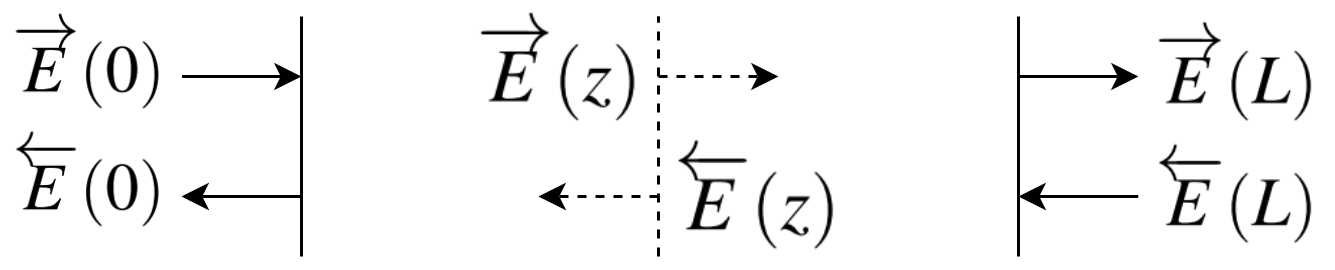}
    \caption{Diagram of the electric fields}
    \label{fig:placeholder}
\end{figure}
In this appendix, we show how to obtain the electric fields inside a Fabry-P\'erot cavity in terms of the fields on the left and right. Introducing transfer matrices $\bm{N}$ and $\bm{M}$, we define
\begin{align}
    \begin{pmatrix}
        \overrightarrow{E}(0) \\
        \overleftarrow{E}(0)  
    \end{pmatrix}    = \bm{N}
    \begin{pmatrix}
       \overrightarrow{E}(z) \\
       \overleftarrow{E}(z) 
    \end{pmatrix} ~, \,
    \begin{pmatrix}
        \overrightarrow{E}(z) \\
        \overleftarrow{E}(z)  
    \end{pmatrix}   = \bm{M}
    \begin{pmatrix}
        \overrightarrow{E}(L) \\
        \overleftarrow{E}(L)  
    \end{pmatrix} 
    ~. 
\end{align}
The forward and backward propagating electric fields can then be calculated at any point inside the cavity, $\overrightarrow{E}(z)$ and $\overleftarrow{E}(z)$, with respect to the known input fields $\overrightarrow{E}(0)$ and $\overleftarrow{E}(L)$:

\begin{align} \label{eq:Ep_at_z}
\begin{aligned}
        \overrightarrow{E}(z) &= \frac{M_{0,0}}{(NM)_{0,0}} \overrightarrow{E}(0) - \frac{M_{0,0} \, (NM)_{0,1}}{(NM)_{0,0}} \overleftarrow{E}(L) + M_{0,1} \overleftarrow{E}(L)  ~, \\
    \overleftarrow{E}(z)  &= \frac{M_{1,0}}{(NM)_{0,0}}\overrightarrow{E}(0) - \frac{M_{1,0} \, (NM)_{0,1}}{(NM)_{0,0}} \overleftarrow{E}(L) + M_{1,1} \overleftarrow{E}(L) ~.
\end{aligned}
\end{align}

\section{Derivation of the Nonlinear Transfer Matrix coefficients for the counter-propagating signal and idler} \label{app:case_a}
In this appendix, we show how to derive the TMM coefficients for the case shown in Fig.~\ref{fig:Beyond_Phase_Matching}a. 
Under the undepleted pump approximation, the coupled differential equations for the left-propagating idler, and the right-propagating signal and pump, are:

\begin{align} \label{eq:contra1}
\begin{aligned}
\frac{d\overrightarrow{A}_s}{dz} &= \frac{2i \omega_s^2 d_{\text{eff}}}{k_s c^2} \overrightarrow{A}_p \overleftarrow{A}_i^* e^{i \Delta k' z}~, \\
\frac{d\overleftarrow{A}_i}{dz} &= -\frac{2i \omega_i^2 d_{\text{eff}}}{k_i c^2} \overrightarrow{A}_p  \overrightarrow{A}_s^* e^{i \Delta k' z}~,
\end{aligned} 
\end{align}
where $\Delta k' = k_p + k_i - k_s$.
For small steps and to first order, we can approximate the solution by integrating, the electric fields become
\begin{align} \label{eq:app:contra_right_Es_Ei_Pr}
\begin{aligned}
    \overrightarrow{E}_s({\scriptstyle L+\Delta L}) &= 
    \,\frac{2\omega_s^2 d_{\text{eff}}}{k_s c^2} \overrightarrow{E}_p({\scriptstyle L}) \overleftarrow{E}_i^*({\scriptstyle L}) 
    \frac{e^{i \Delta k' \, \Delta L } - 1}{\Delta k'}  
    e^{i k_s \Delta L} ~, \\
    \overleftarrow{E}_i({\scriptstyle L}) &= -
    \,\frac{2\omega_i^2 d_{\text{eff}}}{k_i c^2} \overrightarrow{E}_p({\scriptstyle L+\Delta L}) \overrightarrow{E}_s^*({\scriptstyle L+\Delta L}) 
    \frac{e^{-i \Delta k' \, \Delta L } - 1}{\Delta k'}  
    e^{i k_i \Delta L} ~.
\end{aligned}
\end{align}
The transfer matrix coefficients are extracted from Eq.~\ref{eq:app:contra_right_Es_Ei_Pr}
\begin{align}
    w_{14}^f &= \frac{2i\omega_s^2 d_{\text{eff}}}{k_s c^2} \overrightarrow{E}_p({\scriptstyle L}) 
    \Delta L \,  
    \text{sinc}(\Delta k' \, \Delta L /2) e^{i\Delta k' \, \Delta L/2}  
    e^{i k_s \Delta L} ~, \\
    w_{41}^f &= -\frac{2i\omega_i^2 d_{\text{eff}}}{k_i c^2} \overrightarrow{E}_p({\scriptstyle L+ \Delta L}) 
    \Delta L \, 
    \text{sinc}(\Delta k' \, \Delta L /2) e^{-i\Delta k' \, \Delta L/2}   
    e^{i k_i \Delta L} ~.
\end{align}
The coefficients for the other propagation processes (Fig.~\ref{fig:Beyond_Phase_Matching}b-d) can be derived using the same methodology.

\nocite{*}
\bibliography{aipsamp}

@article{franken1961generation,
  title={Generation of optical harmonics},
  author={Franken, eg PA and Hill, Alan E and Peters, CW and Weinreich, Gabriel},
  journal={Physical review letters},
  volume={7},  number={4},  pages={118},
  year={1961},
  publisher={APS}
}

@article{armstrong1962interactions,
  title={Interactions between light waves in a nonlinear dielectric},
  author={Armstrong, John A and Bloembergen, N and Ducuing, J and Pershan, Peter S},
  journal={Physical review},
  volume={127},  number={6},  pages={1918},
  year={1962},
  publisher={APS}
}

@article{giordmaine1965tunable,
  title={Tunable coherent parametric oscillation in LiNb O 3 at optical frequencies},
  author={Giordmaine, Joseph Anthony and Miller, Robert C},
  journal={Physical Review Letters},
  volume={14},  number={24},  pages={973},
  year={1965},
  publisher={APS}
}

@ARTICLE{Stolen1982Parametric,
  author={Stolen, R. and Bjorkholm, J.},
  journal={IEEE Journal of Quantum Electronics}, 
  title={Parametric amplification and frequency conversion in optical fibers}, 
  year={1982},  volume={18},  number={7},
  pages={1062-1072},
  doi={10.1109/JQE.1982.1071660}}

@article{bethune1991optical,
  title={Optical harmonic generation and mixing in multilayer media: extension of optical transfer matrix approach to include anisotropic materials},
  author={Bethune, DS},
  journal={Journal of the Optical Society of America B},
  volume={8},  number={2},  pages={367--373},
  year={1991},
  publisher={Optical Society of America}
}

@article{zelmon1997infrared,
  title={Infrared corrected Sellmeier coefficients for congruently grown lithium niobate and 5 mol.\% magnesium oxide--doped lithium niobate},
  author={Zelmon, David E and Small, David L and Jundt, Dieter},
  journal={Journal of the Optical Society of America B},
  volume={14},  number={12},  pages={3319--3322},
  year={1997},
  publisher={Optical Society of America}
}

@article{sefler1998frequency,
  title={Frequency comb generation by four-wave mixing and the role of fiber dispersion},
  author={Sefler, George A and Kitayama, Ken-ichi},
  journal={Journal of lightwave technology},
  volume={16},  number={9},  pages={1596},
  year={1998},
  publisher={OSA}
}

@article{coen2002supercontinuum,
  title={Supercontinuum generation by stimulated Raman scattering and parametric four-wave mixing in photonic crystal fibers},
  author={Coen, Stephane and Chau, Alvin Hing Lun and Leonhardt, Rainer and Harvey, John D and Knight, Jonathan C and Wadsworth, William J and Russell, Philip St J},
  journal={Journal of the Optical Society of America B},
  volume={19},  number={4},  pages={753--764},
  year={2002},
  publisher={Optical Society of America}
}

@article{lin2003lattice,
  title={Lattice symmetry applied in transfer-matrix methods for photonic crystals},
  author={Lin, Lan-Lan and Li, Zhi-Yuan and Ho, Kai-Ming},
  journal={Journal of Applied Physics},
  volume={94},  number={2},  pages={811--821},
  year={2003},
  publisher={American Institute of Physics}
}

@article{kitaeva2004parametric,
  title={Parametric frequency conversion in layered nonlinear media},
  author={Kitaeva, G Kh and Penin, AN},
  journal={Journal of Experimental and Theoretical Physics},
  volume={98},  number={2},  pages={272--286},
  year={2004},
  publisher={Springer}
}

@article{rohde2005frequency,
  title={Frequency and temporal effects in linear optical quantum computing},
  author={Rohde, Peter P and Ralph, Timothy C},
  journal={Physical Review A—Atomic, Molecular, and Optical Physics},
  volume={71},  number={3},  pages={032320},
  year={2005},
  publisher={APS}
}

@article{yang2008spontaneous,
  title={Spontaneous parametric down-conversion in waveguides: A backward Heisenberg picture approach},
  author={Yang, Zhenshan and Liscidini, Marco and Sipe, John E},
  journal={Physical Review A—Atomic, Molecular, and Optical Physics},
  volume={77},  number={3},  pages={033808},
  year={2008},
  publisher={APS}
}

@book{boyd2008nonlinear,
  title={Nonlinear optics},
  author={Boyd, Robert W and Gaeta, Alexander L and Giese, Enno},
  booktitle={Springer Handbook of Atomic, Molecular, and Optical Physics},
  year={2008},
  publisher={Springer}
}

@article{helt2010spontaneous,
  title={Spontaneous four-wave mixing in microring resonators},
  author={Helt, Lukas G and Yang, Zhenshan and Liscidini, Marco and Sipe, John E},
  journal={Optics letters},
  volume={35},  number={18},  pages={3006--3008},
  year={2010},
  publisher={Optical Society of America}
}

@article{branczyk2010engineered,
  title={Engineered optical nonlinearity for quantum light sources},
  author={Bra{\'n}czyk, Agata M and Fedrizzi, Alessandro and Stace, Thomas M and Ralph, Tim C and White, Andrew G},
  journal={Optics express},  volume={19},  number={1},  pages={55--65},
  year={2010},
  publisher={Optical Society of America}
}

@ARTICLE{Albuquerque2011Transfer,
  author={Albuquerque, André A. C. and Drummond, Miguel V. and Nogueira, Rogério N.},
  journal={Journal of Lightwave Technology}, 
  title={Transfer Matrix and Fourier Transform Methods for Simulation of Second-Order Nonlinear Interactions in a PPLN Waveguide}, 
  year={2011},
  volume={29}, number={24}, pages={3764-3771},
  doi={10.1109/JLT.2011.2174333}}

@article{kauranen2012nonlinear,
  title={Nonlinear plasmonics},
  author={Kauranen, Martti and Zayats, Anatoly V},
  journal={Nature photonics},
  volume={6},  number={11},  pages={737--748},
  year={2012},
  publisher={Nature Publishing Group UK London}
}

@book{nonlinear-fiber-optics,
	title = {Nonlinear Fiber Optics (5th Edition)},
	author = {Agrawal, Govind},
	year = {2013},
	url = {https://app.knovel.com/hotlink/toc/id:kpNFOE0003/nonlinear-fiber-optics/nonlinear-fiber-optics},
	isbn = {978-0-12397-023-7},
	publisher = {Elsevier},
}

@article{liscidini2013stimulated,
  title={Stimulated emission tomography},
  author={Liscidini, Marco and Sipe, JE},
  journal={Physical review letters},
  volume={111},  number={19},  pages={193602},
  year={2013},
  publisher={APS}
}

@article{li2015application,
  title={Application of transfer matrix method to second-harmonic generation in nonlinear photonic bandgap structures: oblique incidence},
  author={Li, Han and Haus, Joseph W and Banerjee, Partha P},
  journal={Journal of the Optical Society of America B},
  volume={32},  number={7},  pages={1456--1462},
  year={2015},
  publisher={Optical Society of America}
}

@article{liu2016resonantly,
  title={Resonantly enhanced second-harmonic generation using III--V semiconductor all-dielectric metasurfaces},
  author={Liu, Sheng and Sinclair, Michael B and Saravi, Sina and Keeler, Gordon A and Yang, Yuanmu and Reno, John and Peake, Gregory M and Setzpfandt, Frank and Staude, Isabelle and Pertsch, Thomas and others},
  journal={Nano letters},
  volume={16},  number={9},  pages={5426--5432},
  year={2016},
  publisher={ACS Publications}
}

@article{huang2018new,
  title={New transfer-matrix method for frequency conversion in nonlinear multilayered structures based on coupled-amplitude equations},
  author={Huang, Jin jer and Feng, Qian and Zhang, Xin Lu and Zhang, Liu Yang},
  journal={Journal of the Optical Society of America B},
  volume={36},  number={1},  pages={26--34},
  year={2018},
  publisher={Optical Society of America}
}

@article{graffitti2018design,
  title={Design considerations for high-purity heralded single-photon sources},
  author={Graffitti, Francesco and Kelly-Massicotte, J{\'e}r{\'e}my and Fedrizzi, Alessandro and Bra{\'n}czyk, Agata M},
  journal={Physical Review A},
  volume={98},  number={5},  pages={053811},
  year={2018},
  publisher={APS}
}

@article{xie2020chip,
  title={On-chip Fabry--P{\'e}rot Bragg grating cavity enhanced four-wave mixing},
  author={Xie, Shengjie and Zhang, Yang and Hu, Yiwen and Veilleux, Sylvain and Dagenais, Mario},
  journal={ACS Photonics},
  volume={7},
  number={4},
  pages={1009--1015},
  year={2020},
  publisher={ACS Publications}
}

@article{zhang2021squeezed,
  title={Squeezed light from a nanophotonic molecule},
  author={Zhang, Y and Menotti, M and Tan, K and Vaidya, VD and Mahler, DH and Helt, LG and Zatti, L and Liscidini, M and Morrison, B and Vernon, Z},
  journal={Nature communications},
  volume={12},  number={1},  pages={2233},
  year={2021},
  publisher={Nature Publishing Group UK London}
}

@book{mackay2022transfer,
  title={The transfer-matrix method in electromagnetics and optics},
  author={Mackay, Tom G and Lakhtakia, Akhlesh},
  year={2022},
  publisher={Springer Nature}
}

@article{luce2022tmm,
  title={TMM-Fast, a transfer matrix computation package for multilayer thin-film optimization: tutorial},
  author={Luce, Alexander and Mahdavi, Ali and Marquardt, Florian and Wankerl, Heribert},
  journal={Journal of the Optical Society of America A},
  volume={39},  number={6},  pages={1007--1013},
  year={2022},
  publisher={Optica Publishing Group}
}

@article{poveda2023custom,
  title={Custom nonlinearity profile for integrated quantum light sources},
  author={Poveda-Hospital, Salvador and Peter, Yves-Alain and Quesada, Nicolas},
  journal={Physical Review Applied},
  volume={19}, number={5}, pages={054033},
  year={2023},
  publisher={APS}
}

@article{sorensen2025simple,
  title={A simple model for entangled photon generation in resonant structures},
  author={Sorensen, Nicholas J and Sultanov, Vitaliy and Chekhova, Maria V},
  journal={Optics Express},
  volume={33},  number={6},  pages={13946--13960},
  year={2025},
  publisher={Optica Publishing Group}
}

@article{PsiQuantum,
    author = {PsiQuantum Team},
    title = {A manufacturable platform for photonic quantum computing},
    journal = Nature,
    year = 2025
}

@article{amores2025generation,
  title={Generation of counterpropagating photon pairs in periodically poled Rb-KTiOPO 4},
  author={Amores, Albert Peralta and Zukauskas, Andrius and Mutter, Patrick and Laurell, Fredrik and Pasiskevicius, Valdas and Swillo, Marcin},
  journal={Physical Review A},
  volume={112},
  number={3},
  pages={032608},
  year={2025},
  publisher={APS}
}

@article{faleo2025optimised,
  title={Optimised spectral purity of unfiltered photons via pump and nonlinearity shaping},
  author={Faleo, Tommaso and Morrison, Christopher L and Beignon, Rom{\'e}o and Graffitti, Francesco and Remesh, Vikas and Frick, Stefan and Fedrizzi, Alessandro and Weihs, Gregor and Keil, Robert},
  journal={arXiv preprint arXiv:2510.06196},
  year={2025}
}

@misc{poveda2025repository,
  author       = {Salvador Poveda-Hospital},
  title        = {TMM applied to Nonlinear Optics},
  year         = {2025},
  howpublished = {\url{https://github.com/salvadorpoveda/TMM_nonlinear}},
  note         = {GitHub repository}
}

\end{document}